\documentclass[aps,reprint]{revtex4-2}
\usepackage{etoolbox} %
\usepackage{blindtext}
\usepackage{graphicx}
\usepackage{subcaption}
\usepackage{amssymb}
\usepackage{xcolor}
\usepackage{graphicx}
\usepackage{floatrow}
\usepackage[fleqn]{amsmath}
\usepackage[font=small,skip=3pt]{caption}
\usepackage{makecell}
\usepackage{hyperref} 
\usepackage{cleveref}
\usepackage{url}
\usepackage{siunitx}

\makeatletter
\appto{\appendix}{%
  \@ifstar{\def\theequation@prefix{A.}}%
          {}%
}
\makeatother

\newcolumntype{B}{>{\color{red}}c}

\begin{document}
\title{Electric Power Enhancement using Spin-Polarized Fuel in Fusion Power Plants}
\author{J. F. Parisi$^{1}$}
\email{jparisi@pppl.gov}
\author{A. Diallo$^1$}
\affiliation{$^1$Princeton Plasma Physics Laboratory, Princeton University, Princeton, NJ, USA}

\begin{abstract}
Using a range of fusion power plant (FPP) concepts, we demonstrate that spin-polarized fuel (SPF) can significantly enhance net electric power output, often by many multiples. Notably, the electric power gain from SPF generally exceeds the corresponding increase in thermal fusion power. Plants close to engineering breakeven stand to benefit most, where even modest boosts in fusion power produce disproportionately larger gains in net electricity. As a representative example, a 25\% increase in fusion power via SPF could allow an ITER-like device (with an added turbine to recover thermal fusion power) to achieve engineering breakeven. These findings strongly motivate the development of spin-polarized fuel for FPPs.
\end{abstract}

\maketitle

\section{Introduction}
Achieving a high net electric power output from fusion power plants (FPPs) requires both increasing the fusion reaction rate and reducing parasitic recirculating power~\cite{Mau1999,Sethian_2005,Najmabadi_2006,Najmabadi2008,Meier2014,Sorbom2015,Menard2016,Federici2019b,Buttery2021,Wade2021,Fradera2021,Schoofs2022,Morris_2022,Menard2023_IAEA,Muldrew2024}. Many FPP designs devote substantial recirculating power to plasma heating, current drive, and other auxiliary systems, such that the total plant output can be comparable to the parasitic load. Consequently, even small multiplicative improvements in thermal fusion power often yield large gains in net electric power. In this work, we show that spin-polarized fuel (SPF) offers a promising pathway to amplify net electric power in a range of confinement concepts, even in their presently optimized plasma scenarios.

SPF can enhance the fusion cross section by polarizing the nuclear spins of deuterium-tritium (DT) fuel~\cite{Kulsrud1986,Ciullo_2016,Sandorfi2017}. Under ideal conditions, fully aligned DT spins can increase the DT fusion cross section by as much as 50\%. Additional effects, such as higher alpha heating, may further raise the plasma temperature and fusion reactivity, yielding a total fusion power increase of up to 80--90\% in certain regimes~\cite{Bittoni1983,Smith2018IAEA,Heidbrink2024}. A recent analysis~\cite{Meneghini2024,Heidbrink2024} showed that using SPF to achieve a 50\% boost in thermal fusion power can translate into a 90\% increase in net electric output for a compact FPP~\cite{Weisberg2023}. This disproportionate enhancement arises whenever the FPP exceeds engineering breakeven.

Building on these results, in \Cref{sec:powerincrease} we show how employing SPF can amplify the net electric power $P_{\mathrm{e}}$ \textit{by a significantly larger factor than the underlying gain in fusion power}. The resulting increase in net electricity can be an order of magnitude or more when FPPs operate near engineering breakeven. This outcome underscores the significant potential of spin-polarized fuel to accelerate progress toward commercially viable fusion. We provide a summary and discuss future steps in \Cref{sec:discussion}.

\section{Electric Power Gain} \label{sec:powerincrease}

\begin{figure}[tb!]
    \centering
    \begin{subfigure}[t]{0.98\textwidth}
    \centering
    \includegraphics[width=1.0\textwidth]{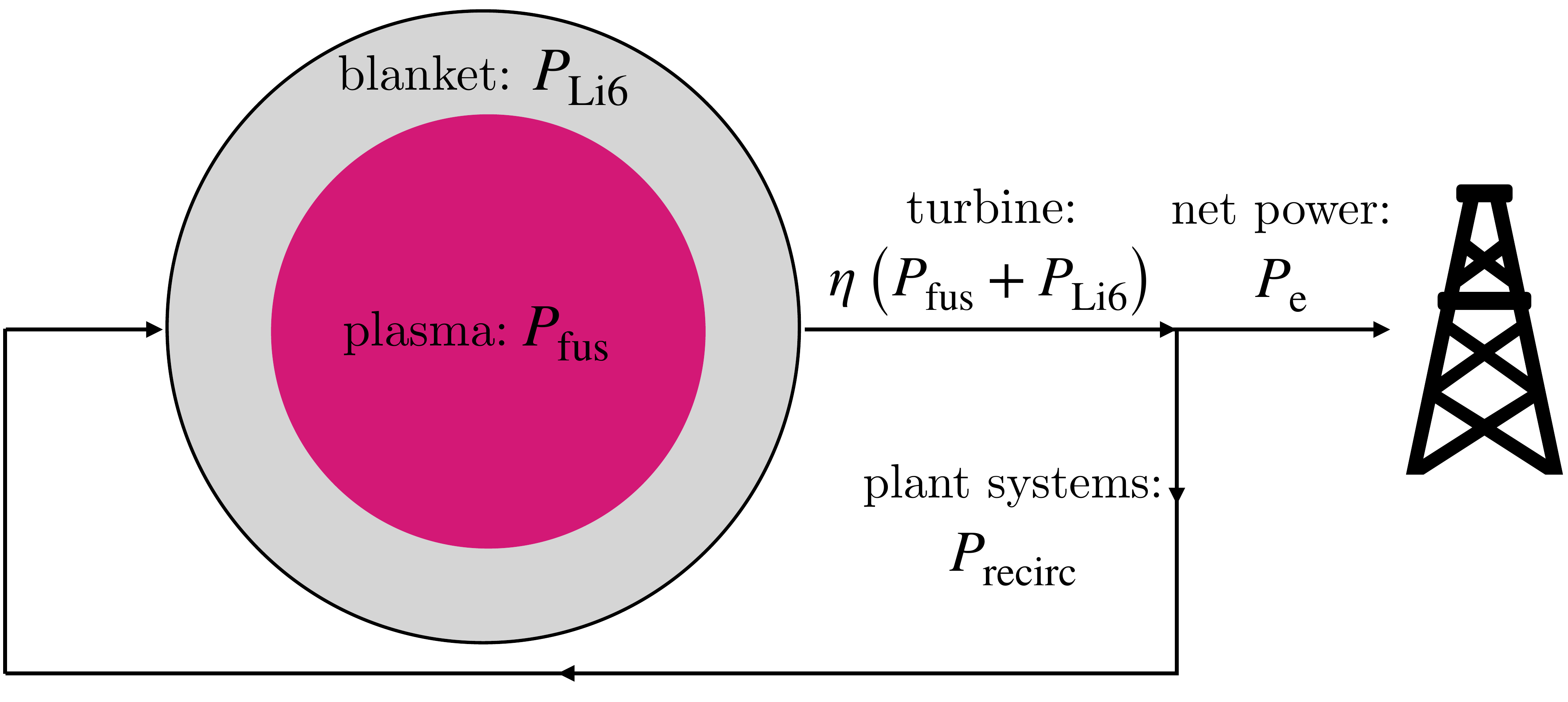}
    \end{subfigure}
    \caption{Simplified power model used in this work.}
    \label{fig:powerschematic}
\end{figure}

A simple way to model the net electric power output $P_\mathrm{e}$ of a FPP -- illustrated schematically in \Cref{fig:powerschematic} -- is
\begin{equation}
    P_\mathrm{e} = \eta P_\mathrm{th} - P_\mathrm{recirc},
    \label{eq:Pe_notritium}
\end{equation}
where $P_\mathrm{th}$ is the total thermal power, $\eta$ is the thermodynamic conversion efficiency, and $P_\mathrm{recirc}$ is the recirculating power. The total thermal power arises from both the fusion power $P_\mathrm{fus}$ and the exothermic reaction power due to lithium-6 neutron capture (for simplicity we assume 100\% Li6) $P_{\mathrm{Li}6}$:
\begin{equation}
    P_\mathrm{th} = P_\mathrm{fus} + P_{\mathrm{Li}6} = \chi P_\mathrm{fus},
\end{equation}
where
\begin{equation}
\chi \equiv 1 + \mathrm{TBR}\,\frac{E_{\mathrm{Li}6}}{E_{\mathrm{fus}}}.
\label{eq:chi}
\end{equation}
Here, the tritium breeding ratio (TBR) is taken to be 1.1, assuming that all tritium is bred via the exothermic reaction with lithium-6 ($E_{\mathrm{Li}6} = 4.8~\mathrm{MeV}$). The DT fusion reaction releases $E_{\mathrm{fus}} = 17.6~\mathrm{MeV}$. To account for the effects of SPF, we introduce a multiplicative factor $\mathcal{P}$:
\begin{equation}
    P_\mathrm{th} = \mathcal{P} P_\mathrm{th,base},
    \label{eq:Pfusion_polarization}
\end{equation}
where $\mathcal{P} > 1$ indicates the fractional increase in both the base fusion power $P_\mathrm{fus,base}$ and the base lithium-6 power $P_{\mathrm{Li}6,base}$. “Base” quantities refer to the case without polarization ($\mathcal{P} = 1$). Because $\mathcal{P} P_\mathrm{fus,base}$ implies a higher neutron production rate, $P_{\mathrm{Li}6}$ also scales accordingly. For ARC-class and ARIES-CS designs, instead of computing $\chi$ via \Cref{eq:chi}, we use values directly reported in their respective design references \cite{Sorbom2015,Najmabadi2008}.

\begin{figure}[bt!]
    \centering
    \begin{subfigure}[t]{0.98\textwidth}
    \centering
    \includegraphics[width=1.0\textwidth]{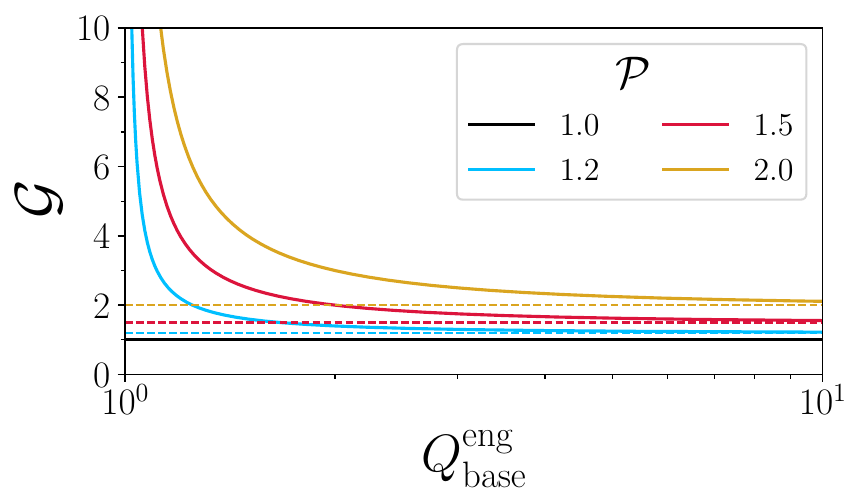}
    \end{subfigure}
    \caption{Net electric power enhancement factor $\mathcal{G}$ (\Cref{eq:Peratio}) plotted against the engineering gain for unpolarized fuel, $Q^\mathrm{eng}_\mathrm{base}$. Four different values for the fusion power multiplier $\mathcal{P}$ (\Cref{eq:Pfusion_polarization}) are shown. Dashed lines highlight constant $\mathcal{P}$ values.}
    \label{fig:Gsimplie}
\end{figure}

To measure how SPF enhances net electric power, we define
\begin{equation}
    \mathcal{G} \equiv \frac{P_\mathrm{e}}{|P_\mathrm{e,base}|} \;=\; \frac{\mathcal{P}\,Q^\mathrm{eng}_\mathrm{base} \;-\; 1}{\bigl|\,Q^\mathrm{eng}_\mathrm{base} - 1\,\bigr|},
    \label{eq:Peratio}
\end{equation}
where $Q^\mathrm{eng}$ is the {engineering gain}:
\begin{equation}
    Q^\mathrm{eng} \equiv \frac{\eta\,P_\mathrm{th}}{P_\mathrm{recirc}}.
    \label{eq:Qeng}
\end{equation}
A net production of electricity requires $Q^\mathrm{eng} > 1$. \Cref{fig:Gsimplie} illustrates how $\mathcal{G}$ behaves as a function of $Q^\mathrm{eng}_\mathrm{base}$ for four different values of $\mathcal{P}$. In every case, $\mathcal{G}$ exceeds $\mathcal{P}$ when $\mathcal{P} > 1$. Moreover, as $Q^\mathrm{eng}_\mathrm{base} \to 1^+$, $\mathcal{G} \to \infty$. We now proceed to apply this argument to several fusion power plant concepts.

\subsection{STEP-Class}
We first apply these ideas to a Spherical Tokamak for Energy Production (STEP)-class design~\cite{Fradera2021,Schoofs2022,Tholerus_2024}, considering two hypothetical scenarios. STEP1 assumes $P_\mathrm{fus} = 1800$~MW with a base net power $P_\mathrm{e,base} = 100$~MW, whereas STEP2 uses $P_\mathrm{fus} = 1600$~MW with $P_\mathrm{e,base} = 250$~MW. Key parameters are summarized in \Cref{table:PowerCases}. In both scenarios, we scan three possible thermodynamic conversion efficiency values, $\eta = 0.3,\,0.4,\,0.5$, and determine $P_\mathrm{e}$ by adjusting $P_\mathrm{recirc}$ accordingly.

\begin{figure}[bt]
    \centering
    \begin{subfigure}[t]{0.94\textwidth}
    \centering
    \includegraphics[width=1.0\textwidth]{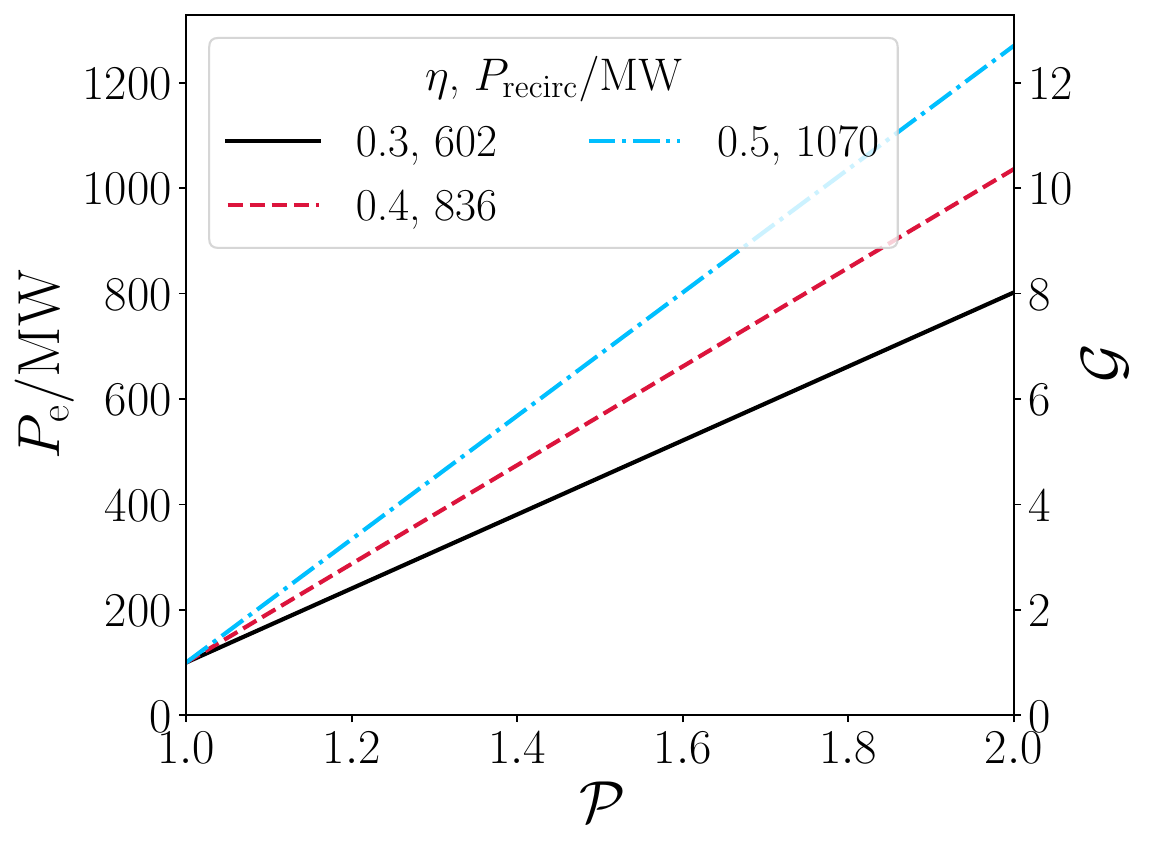}
    \caption{$P_\mathrm{e}$ and $\mathcal{G}$ for STEP1}
    \end{subfigure}
    \begin{subfigure}[t]{0.94\textwidth}
    \centering
    \includegraphics[width=1.0\textwidth]{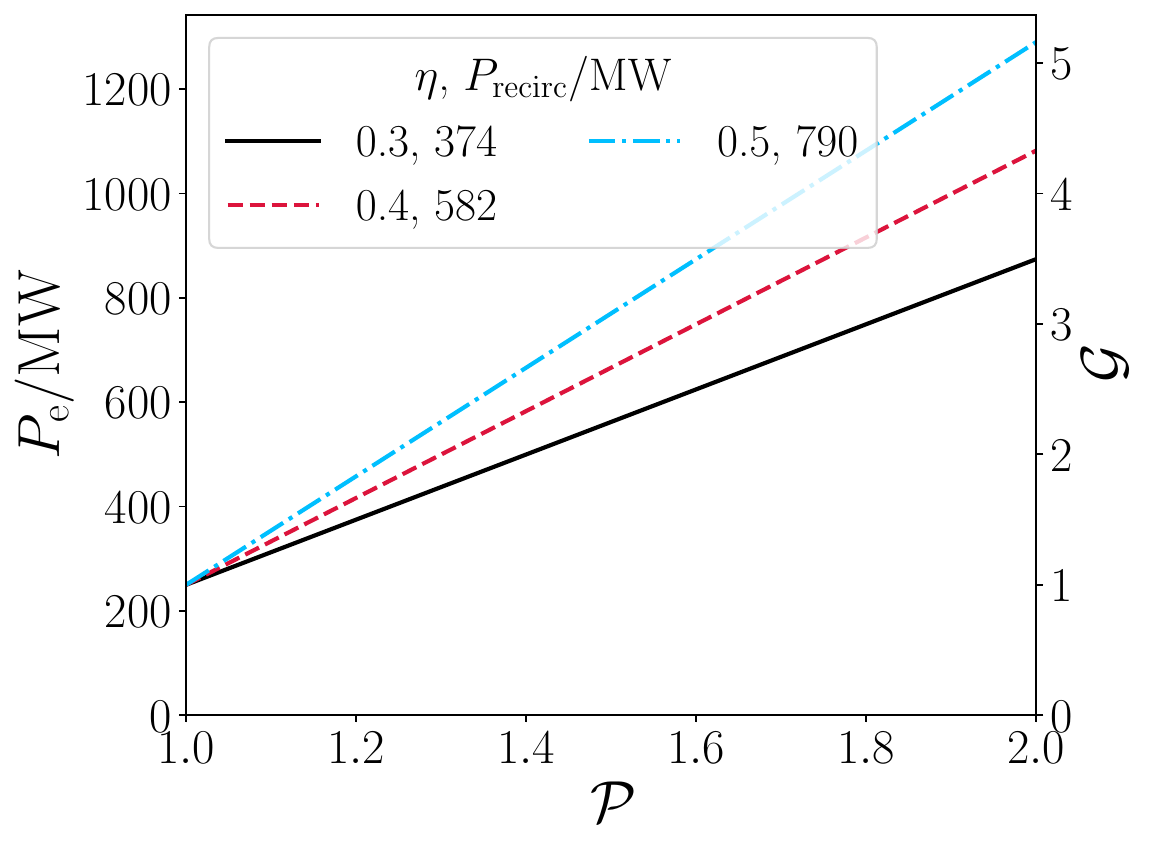}
    \caption{$P_\mathrm{e}$ and $\mathcal{G}$ for STEP2}
    \end{subfigure}
    \caption{Net electric power $P_\mathrm{e}$ versus polarization power increase factor $\mathcal{P}$ for (a) STEP1 and (b) STEP2. The right $y$-axis is $\mathcal{G} \equiv P_\mathrm{e}/|P_\mathrm{e,base}|$ (\Cref{eq:Peratio}).}
    \label{fig:electric_output_STEP}
\end{figure}

\Cref{fig:electric_output_STEP}(a) and (b) show how $P_\mathrm{e}$ depends on $\mathcal{P}$ for STEP1 and STEP2. In STEP1, for instance, unpolarized fuel yields $P_\mathrm{e} = 100$~MW. By polarizing the fuel enough to achieve $\mathcal{P} = 1.9$, $P_\mathrm{e}$ can climb to $732$~MW at $\eta = 0.3$ or $1153$~MW at $\eta = 0.5$, corresponding to net power gains of 7.3 and 11.5 times the base value (unpolarized case with $\mathcal{P} = 1$). A more modest polarizaton of $\mathcal{P} = 1.5$ still boosts $P_\mathrm{e}$ by factors of 4.5 and 6.9 for $\eta=0.3$ and $\eta = 0.5$. STEP2 exhibits similar trends but yields smaller enhancements, since it begins farther above the threshold $Q^\mathrm{eng}_\mathrm{base} = 1$. These polarized scenarios are labeled STEP1-P and STEP2-P in \Cref{table:PowerCases}.

\subsection{ARC-Class}
\begin{figure}[bt!]
    \centering
    \begin{subfigure}[t]{0.85\textwidth}
    \centering
    \includegraphics[width=1.0\textwidth]{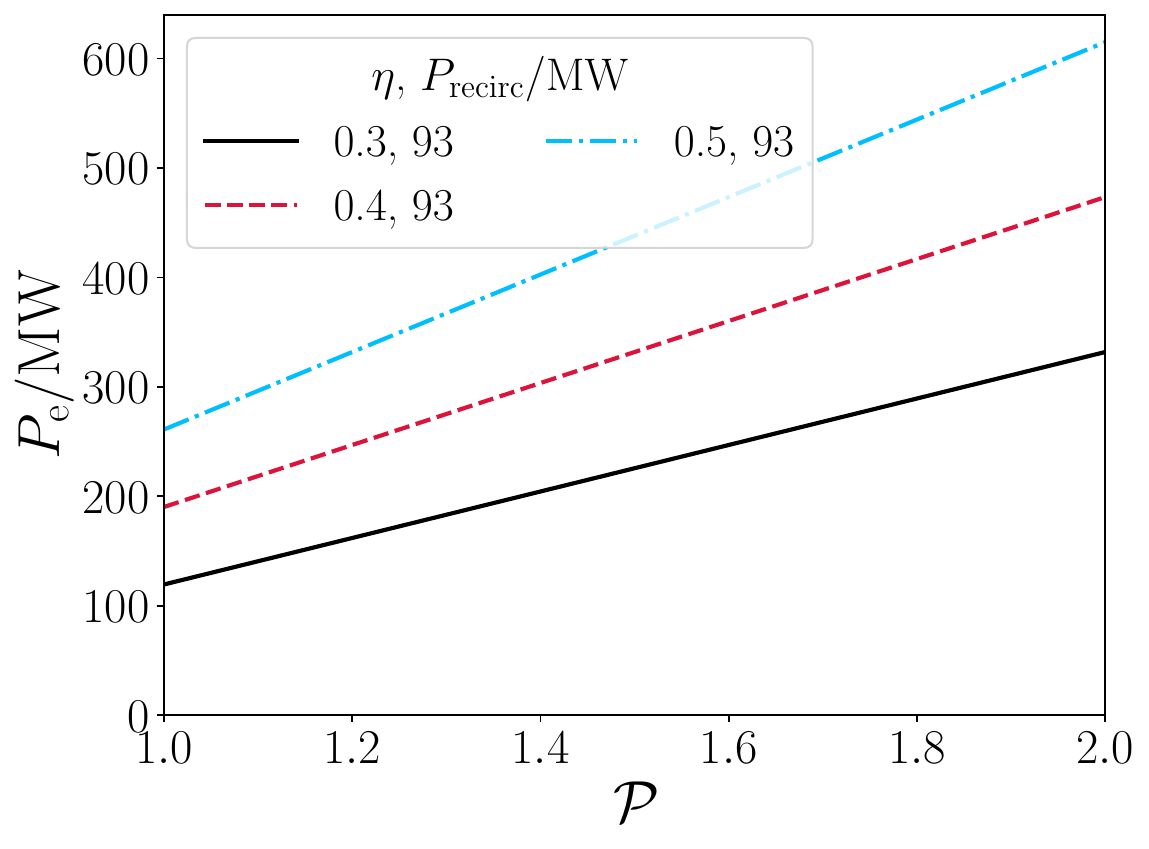}
    \caption{$P_\mathrm{e}$}
    \end{subfigure}
    \centering
    \begin{subfigure}[t]{0.98\textwidth}
    \centering
    \includegraphics[width=1.0\textwidth]{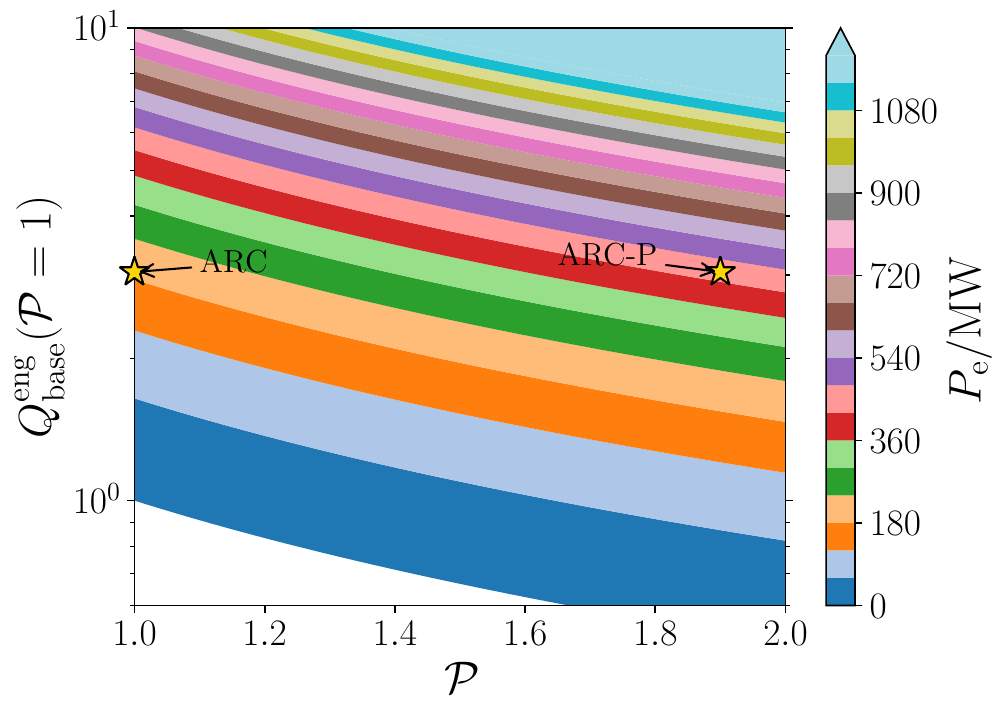}
    \caption{$Q^\mathrm{eng}_\mathrm{base}$ and $P_\mathrm{e}$}
    \end{subfigure}
    \caption{(a) Net electric power $P_\mathrm{e}$ versus polarization power increase factor $\mathcal{P}$ for ARC-class at three different conversion efficiencies $\eta$. (b) Contours of $P_\mathrm{e}$ versus $\mathcal{P}$ and $Q^\mathrm{eng}_\mathrm{base}$ at fixed $P_\mathrm{recirc}=93$~MW and $\eta=0.4$. The nominal ARC parameters are indicated.}
    \label{fig:electric_output_ARC}
\end{figure}

A similar approach applies to the ARC-class FPP design~\cite{Sorbom2015}. Here, we fix $P_\mathrm{recirc} = 93$~MW (deduced from~\cite{Sorbom2015}) and scan $\eta$. \Cref{fig:electric_output_ARC}(a) shows $P_\mathrm{e}$ versus $\mathcal{P}$ for three choices of $\eta$. Although base ARC has $Q^\mathrm{eng}_\mathrm{base} = 3.05$, adding SPF still approximately doubles the net electric power. For instance, at $\eta = 0.4$, $P_\mathrm{e}$ rises from 190~MW to 445~MW as $\mathcal{P}$ reaches 1.9, corresponding to $\mathcal{G} = 2.34$. \Cref{fig:electric_output_ARC}(b) illustrates how $P_\mathrm{e}$ depends on both $\mathcal{P}$ and $Q^\mathrm{eng}_\mathrm{base}$, emphasizing that gains become largest when $Q^\mathrm{eng}_\mathrm{base}$ is near unity.

\subsection{ITER-like}

\begin{figure}[bt!]
    \centering
    \begin{subfigure}[t]{0.98\textwidth}
    \centering
    \includegraphics[width=1.0\textwidth]{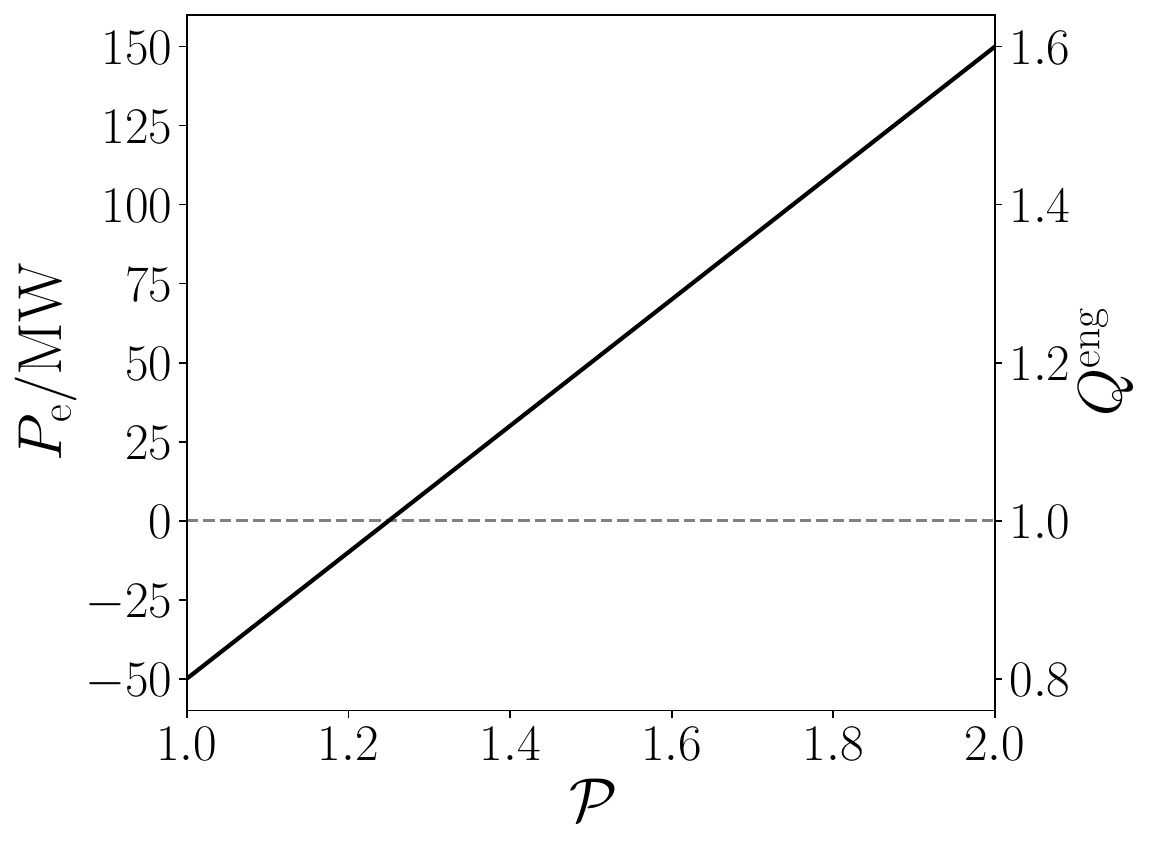}
    \caption{$P_\mathrm{e}$ and $Q^\mathrm{eng}$}
    \end{subfigure}
    \centering
    \begin{subfigure}[t]{0.98\textwidth}
    \centering
    \includegraphics[width=1.0\textwidth]{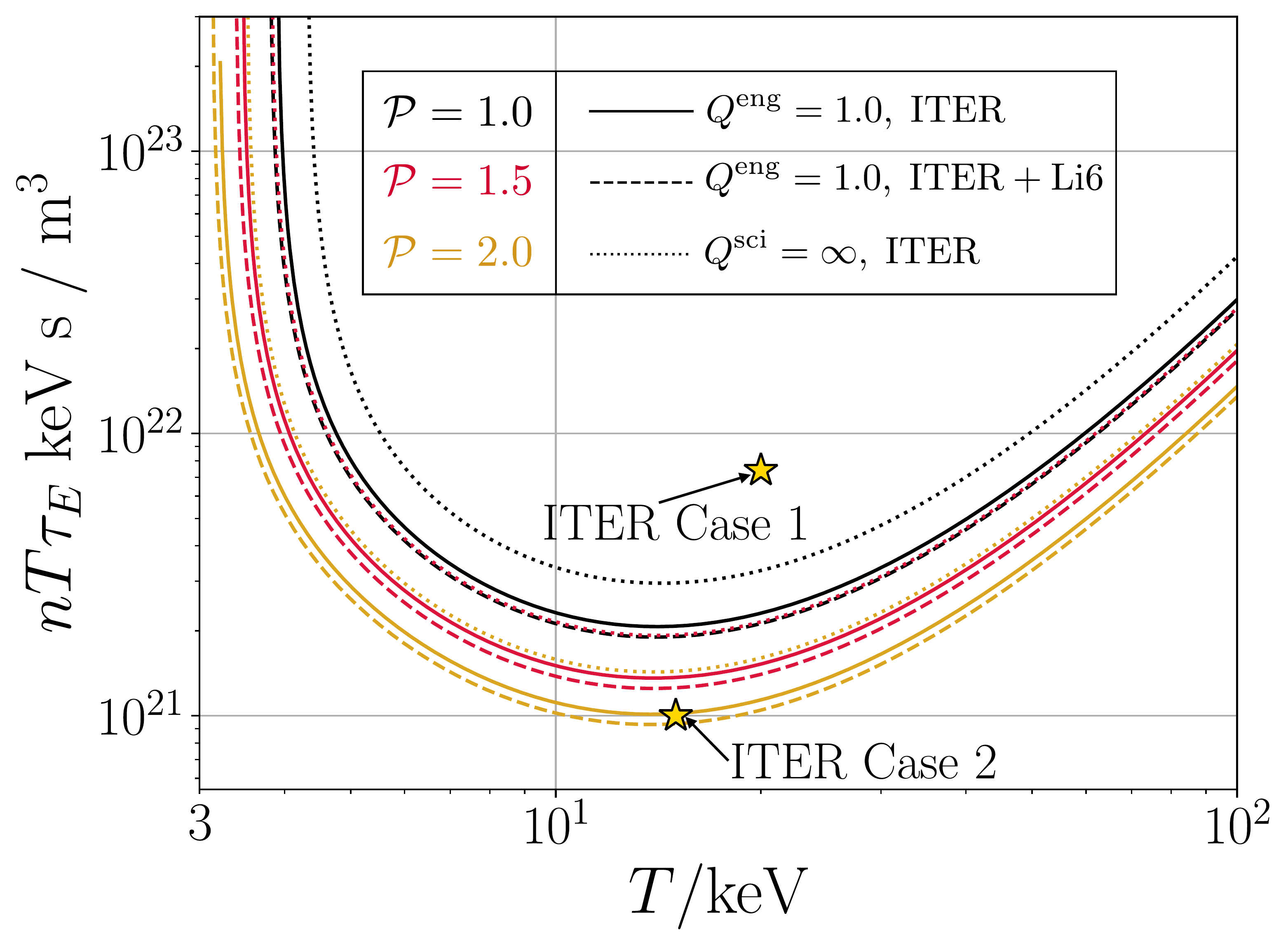}
    \caption{$n T \tau_E$}
    \end{subfigure}
    \caption{(a) Net electric power $P_\mathrm{e}$ and engineering gain $Q^\mathrm{eng}$ (right $y$-axis) versus polarization multiplier $\mathcal{P}$ for an ITER-like device without a Li-6 breeder. (b) The required triple product $nT\tau_E$ for $Q^\mathrm{eng}=1$ as a function of $T$ (\Cref{eq:tripleproduct_engineering_main}), showing how increasing $\mathcal{P}$ lowers the triple product requirement. ITER-like Case 1 and Case 2 represent two different projections for ITER’s steady-state operation. }
    \label{fig:electric_output_ITER}
\end{figure}

We now study the effect of SPF in an ITER-like device, which unlike the current machine \cite{Iter1999,Shimada2007}, is equipped to convert heat into electricity in two configurations: (i) `ITER' that captures thermal fusion energy and converts into electricity with an efficiency of $\eta = 0.4$, and (ii) `ITER+Li6' that also captures energy from Li6 neutron capture with TBR = 1.1. \Cref{fig:electric_output_ITER}(a) shows how $P_\mathrm{e}$ for ITER and ITER+Li6 varies with $\mathcal{P}$, with the right $y$-axis indicating $Q^\mathrm{eng}$. Even modest $\mathcal{P}$ shifts ITER’s operating point closer to net electricity production if one recovers the thermal fusion power with a turbine -- note that we have neglected the exothermic Li-6 reactions. 

We next show how SPF reduces the triple product required for engineering breakeven--\(Q^\mathrm{eng} = 1\)--in an ITER-like device. The triple product \(nT\,\tau_E\) \cite{Lawson1957,Wurzel2022} is a useful figure of merit, combining the sum of the D and T densities \(n\), the plasma temperature \(T\), and energy confinement time \(\tau_E\). The standard form of $nT\tau_E$ describes the conditions for achieving a self-sustaining fusion reaction by balancing energy generation and loss mechanisms. A lower required $nT\,\tau_E$ indicates that the plasma can reach net energy production under less demanding conditions. We extend the calculation to describe the triple product required for engineering breakeven.

In \Cref{app:triple_product_engineering} we derive the triple product for engineering breakeven,
\begin{equation}
nT\,\tau_E \bigl|_{Q^\mathrm{eng} } \;=\;
\frac{12\,T^2}{\bigl(f + \delta\,\eta\,\chi \bigr)\,\mathcal{P}\,\langle \sigma v\rangle\,E_{\mathrm{fus}} \;-\; 4\,C_B\,T^{1/2}}.
\label{eq:tripleproduct_engineering_main}
\end{equation}
Here, \(f \simeq 0.2\) represents the fraction of fusion power carried by alpha particles, \(\delta \equiv P_\mathrm{ext}/P_\mathrm{recirc}\) denotes the fraction of recirculating power absorbed by the plasma from external heating with power $P_\mathrm{ext}$, and \(C_B\) captures Bremsstrahlung losses. \Cref{fig:electric_output_ITER}(b) illustrates how higher \(\mathcal{P}\) values (associated with SPF) reduce the required triple product for \(Q^\mathrm{eng}=1\), indicating that SPF can substantially ease the path to engineering breakeven. In \Cref{fig:electric_output_ITER}(b) we also consider ITER+Li6, which has a lower required triple product than ITER. Additionally, we plot contours for ignited plasmas with $Q^\mathrm{sci}\equiv P_\mathrm{fus}/P_\mathrm{ext} \to \infty$. Consistent with \cite{Mitarai1992}, increasing $\mathcal{P}$ decreases the required triple product for ignition. The `ITER Case 1' marker is based on~\cite{Wagner2009,Wurzel2022} and the `ITER Case 2' marker is a degraded hypothetical scenario.

From a mathematical standpoint, boosting $\mathcal{P}$ resembles improving the confinement enhancement factor $H$ for energy confinement \cite{Goldston1984,Kaye1985}, where $\tau_E = H\,\langle\tau_E\rangle$ for the nominal $\langle\tau_E\rangle$ confinement, as long as radiative losses remain small. Thus, just as an elevated $H$ can drastically reduce overnight capital costs~\cite{Wade2021}, a higher $\mathcal{P}$ from SPF could have a similarly strong impact.

\begin{figure}[!]
    \centering
    \begin{subfigure}[t]{0.88\textwidth}
    \centering
    \includegraphics[width=1.0\textwidth]{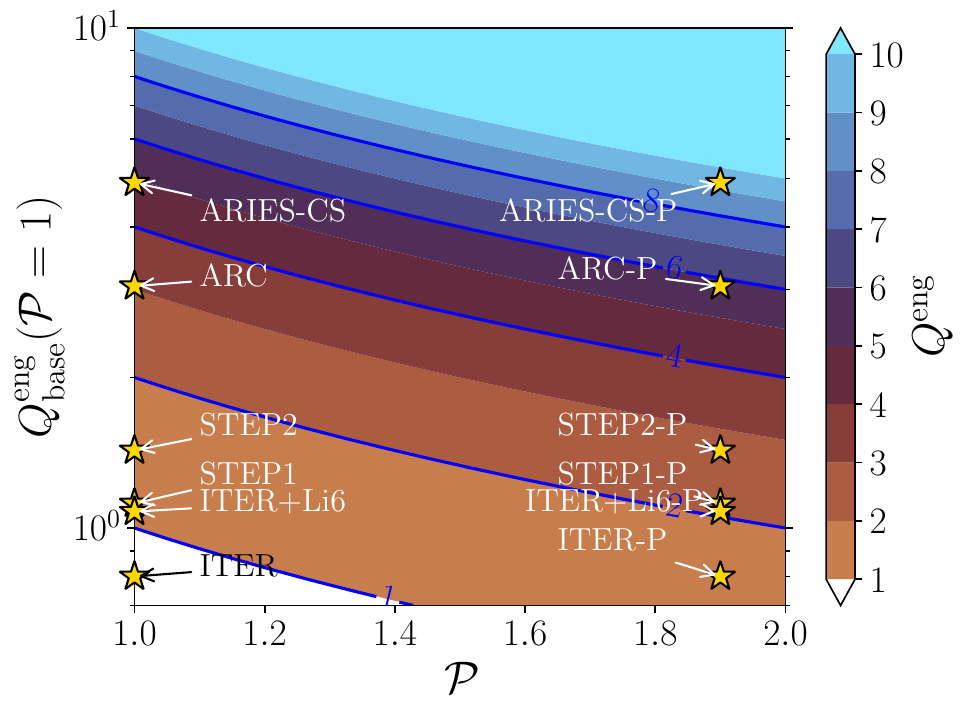}
    \caption{$Q^\mathrm{eng}_\mathrm{base}$ and $Q^{\mathrm{eng}}$}
    \end{subfigure}
    \centering
    \begin{subfigure}[t]{0.88\textwidth}
    \centering
    \includegraphics[width=1.0\textwidth]{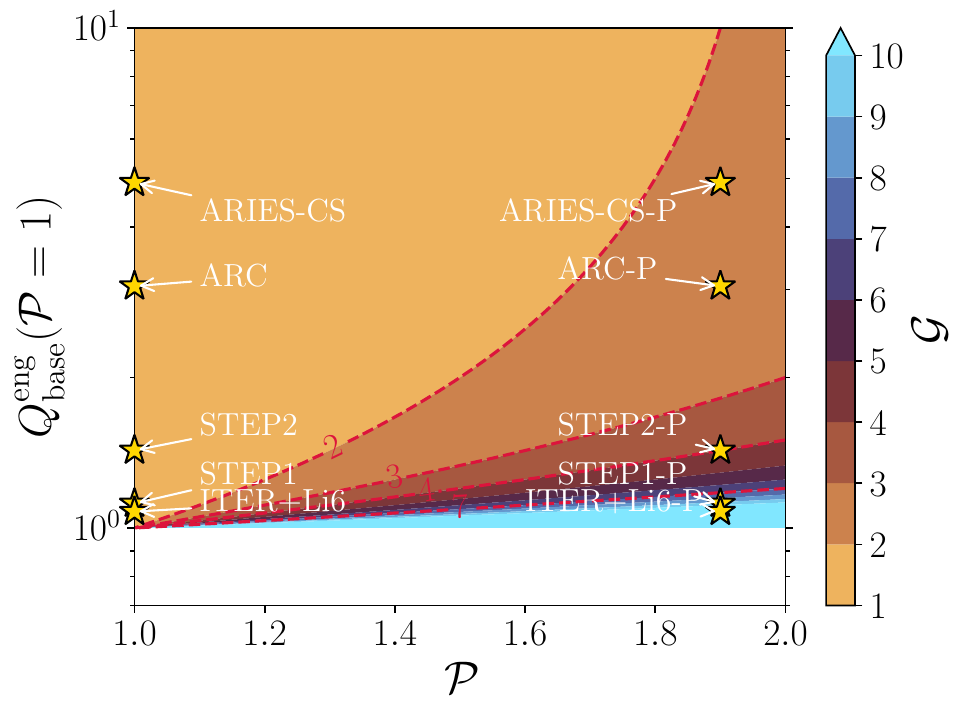}
    \caption{$Q^\mathrm{eng}_\mathrm{base}$ and $\mathcal{G}$}
    \end{subfigure}
    \centering
    \begin{subfigure}[t]{0.88\textwidth}
    \centering
    \includegraphics[width=1.0\textwidth]{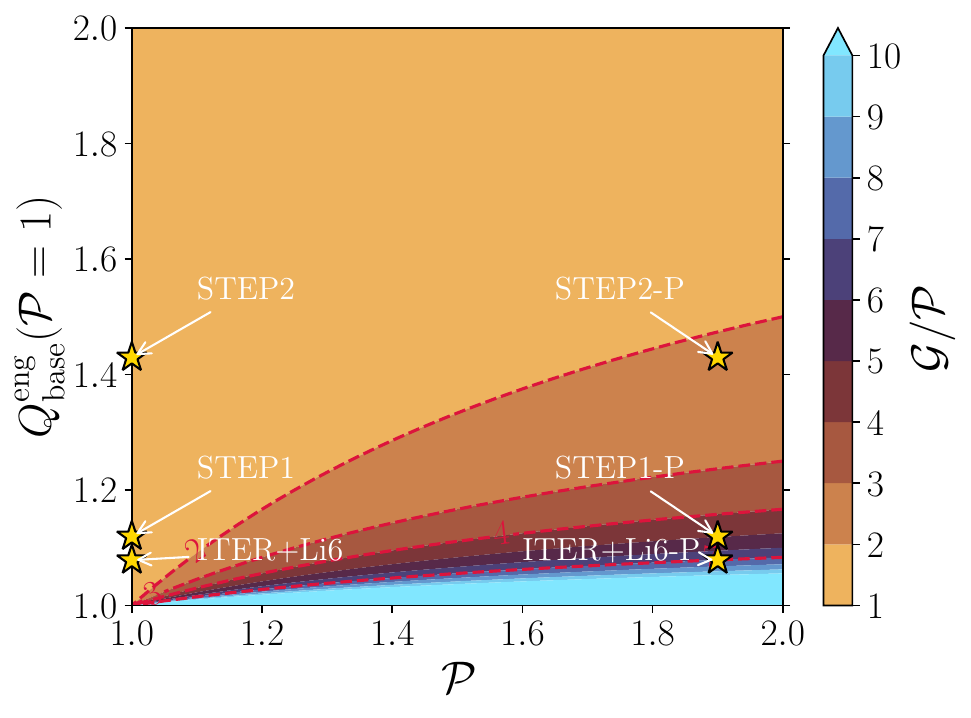}
    \caption{$Q^\mathrm{eng}_\mathrm{base}$ and $\mathcal{G}/\mathcal{P}$}
    \end{subfigure}
    \caption{(a) Engineering gain $Q^{\mathrm{eng}}$, (b) net electric power enhancement $\mathcal{G}$, (c) a zoomed view of the ratio $\mathcal{G}/\mathcal{P}$ for $1 \leq Q^\mathrm{eng}_\mathrm{base} \leq 2$. All are plotted as functions of $\mathcal{P}$ and $Q^\mathrm{eng}_\mathrm{base}$.}
    \label{fig:pelectric_general}
\end{figure}

\subsection{ARIES-CS}
We also consider the ARIES-CS compact stellarator \cite{Najmabadi2008}, which features a lower recirculating power fraction than the tokamak designs discussed above. Taking $P_{\mathrm{Li}6} = 480$~MW and $\eta = 0.43$ from the design study, \Cref{table:PowerCases} shows that increasing $\mathcal{P}$ from 1 to 1.9 raises $P_\mathrm{e}$ from $1000$~MW to $2130$~MW. Although $Q^\mathrm{eng}_\mathrm{base}=4.90$ is relatively high, doubling the net electric power is still a substantial benefit.

\subsection{Multi-Device Comparison}
\Cref{table:PowerCases} compares base (unpolarized) and polarized scenarios for STEP1, STEP2, ARC, ITER+Li6, ITER, and ARIES-CS. In \Cref{fig:pelectric_general}(a), $Q^\mathrm{eng}$ grows roughly linearly with $\mathcal{P}$ as long as $Q^\mathrm{eng}_\mathrm{base} > 1$. However, \Cref{fig:pelectric_general}(b) shows that $\mathcal{G}$ (the net power enhancement) increases more strongly in a {nonlinear} fashion. Designs that start near $Q^\mathrm{eng}_\mathrm{base}=1$, such as STEP1, STEP2, or ITER+Li6, see the largest jumps in net electric power. 

Another useful metric is $\mathcal{G}/\mathcal{P}$, shown in \Cref{fig:pelectric_general}(c). For designs that begin only slightly above breakeven, $\mathcal{G}/\mathcal{P}$ can be much larger than unity. In the ITER+Li6 example, $\mathcal{G}/\mathcal{P}=6.89$, indicating the net electric power multiplier is nearly seven times bigger than the direct fusion power multiplier $\mathcal{P}$. This underscores the disproportionate benefits SPF can confer on systems that are borderline in terms of net power generation.

\begin{figure*}[tb!]
    \centering
    \begin{subfigure}[t]{0.48\textwidth}
    \centering
    \includegraphics[width=1.0\textwidth]{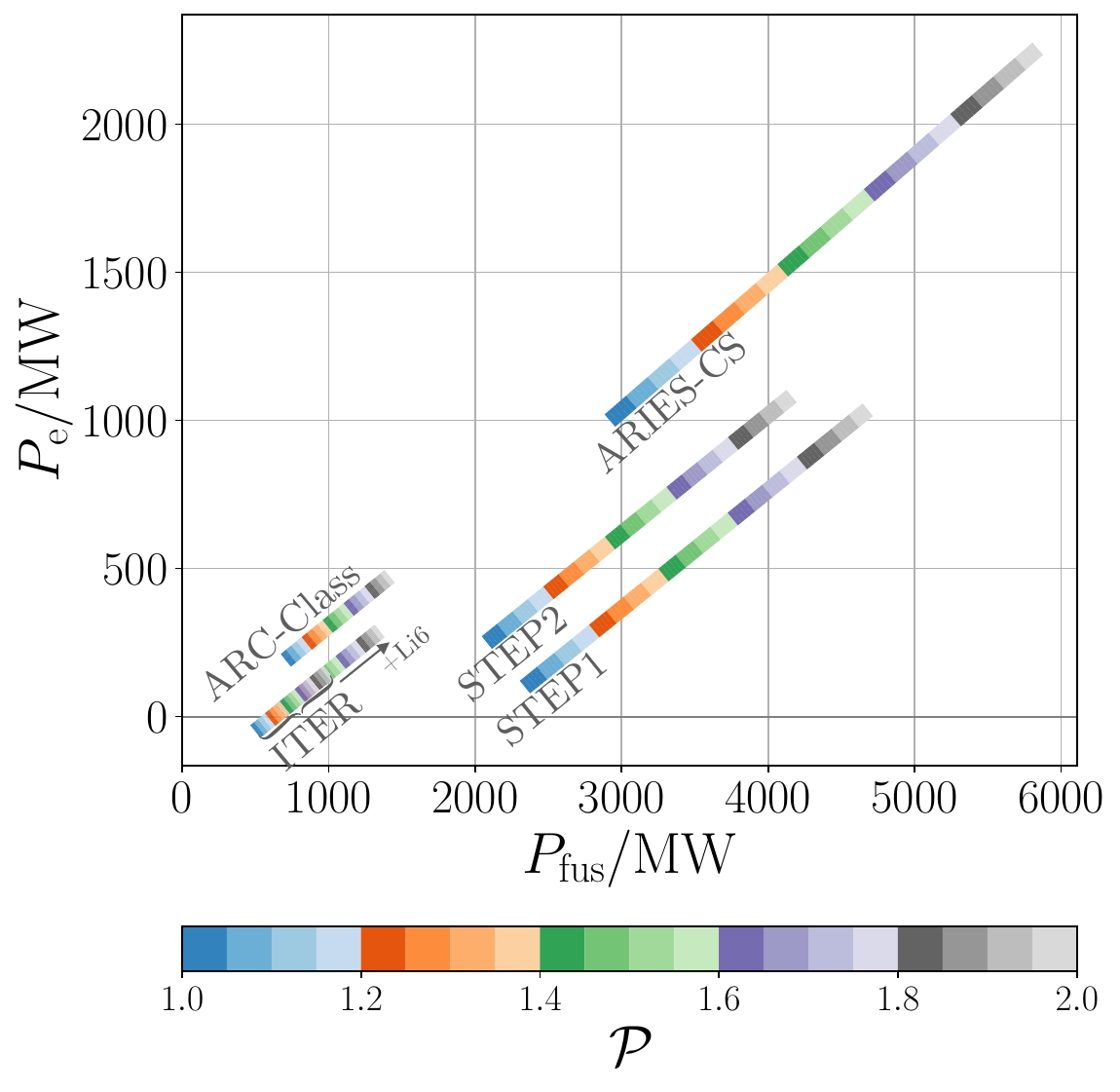}
    \caption{}
    \end{subfigure}
    \centering
    \begin{subfigure}[t]{0.47\textwidth}
    \centering
    \includegraphics[width=1.0\textwidth]{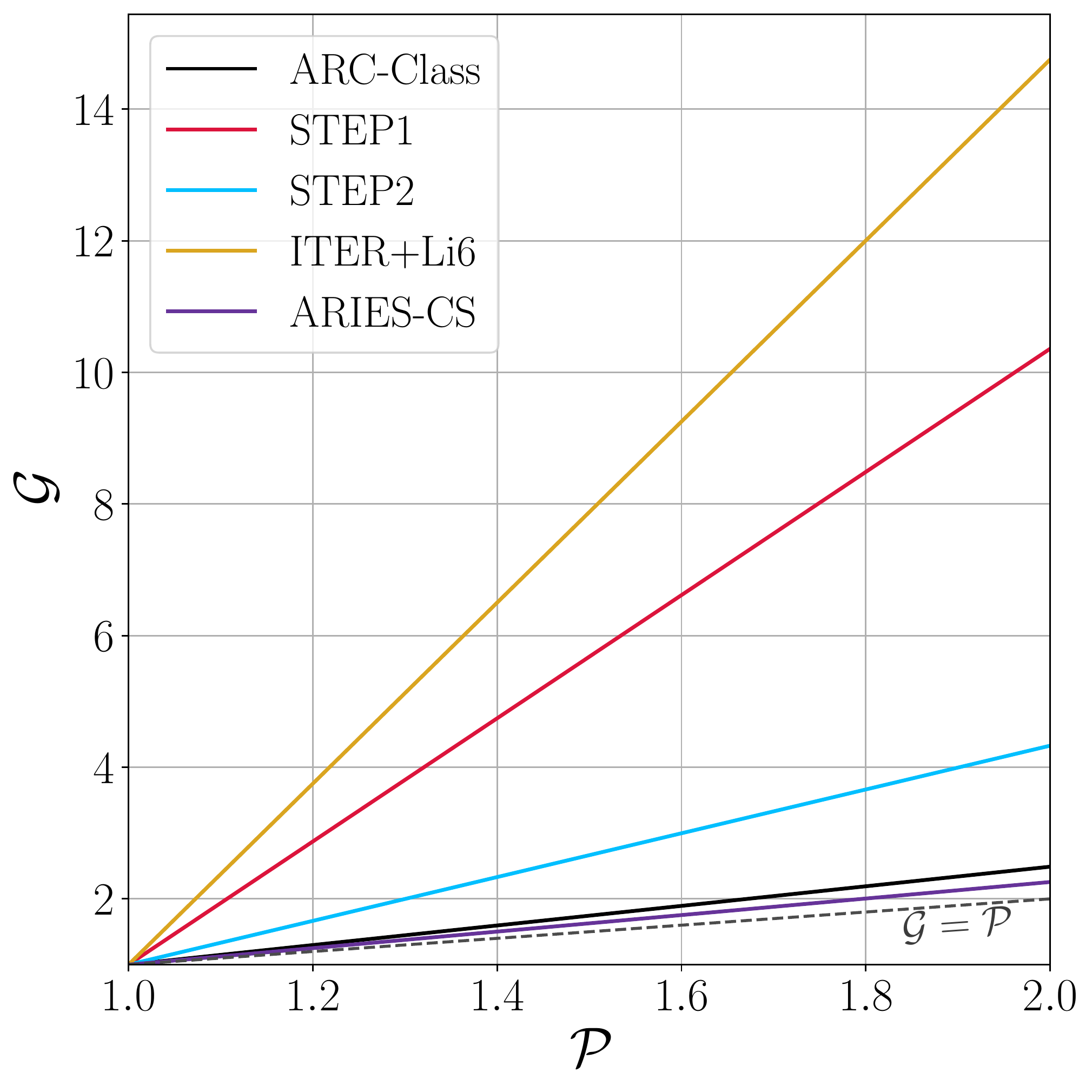}
    \caption{}
    \end{subfigure}    \caption{(a) net electric power $P_\mathrm{e}$ versus fusion power $P_\mathrm{fus}$ for different $\mathcal{P}$ values, (b) Net electric gain $\mathcal{G}$ versus $\mathcal{P}$. Grey-dashed line indicates $\mathcal{G}=\mathcal{P}$.}
    \label{fig:summary}
\end{figure*}

\section{Discussion} \label{sec:discussion}

Building on a recent analysis~\cite{Heidbrink2024}, we have demonstrated that using spin-polarized fuel (SPF) can enhance the net electric power output of fusion power plants (FPPs) by at least as much as the increase in fusion reactivity. In systems near engineering breakeven, a modest rise in fusion power can yield a disproportionately larger boost in net electric power, helping borderline FPPs move from breakeven to substantial net production. Because commercial viability depends on electricity output rather than raw fusion power, improvements that increase net output—especially in plants with high recirculating power fractions and lower thermodynamic efficiencies—offer a near-term route to compensate for limited gains in conventional efficiency enhancements.

\Cref{fig:summary} summarizes the main results for machines studied in this work. \Cref{fig:summary}(a) shows the net electric power and thermal fusion power as a function of $\mathcal{P}$ for different machines. \Cref{fig:summary}(b) shows electric gain $\mathcal{G}$ versus polarization multiplier $\mathcal{P}$ -- for two machines $\mathcal{G}$ exceeds 10 at high $\mathcal{P}$ values. For all the machines, $\mathcal{G} > \mathcal{P}$, showing the net electric power is amplified more strongly than the fusion power.

This paper presents a simplified model rather than a detailed engineering design. Many challenges remain. For example, power exhaust solutions such as liquid metals~\cite{Coenen2014,Tabares2015,Van2017,Allain2019,Andruczyk2020,Berkery_2024} and highly radiative scenarios \cite{Iter1999_powerparticle,Takenaga2005,Loarte2011,Casali2025} must be developed to handle higher steady-state heat fluxes, and although SPF fueling systems are not expected to draw significantly more power~\cite{Baylor2023}, additional costs such as increased blanket pumping and cooling need evaluation. Higher neutron wall loads~\cite{Kulsrud1986,Bae_2025} may exacerbate already challenging material degradation~\cite{Zinkle2009,Knaster2016,Bhattacharya2022,El2023,Tong2025}, and it remains uncertain whether SPF retains sufficient polarization in harsh fusion environments~\cite{Kulsrud1986,Didelez2011,Garcia2023,Hu2023,Heidbrink2024,Garcia2025}, although a recent result is promising~\cite{Zheng2024}. Furthermore, scaling up high-throughput, high-polarization sources is an ongoing technical challenge~\cite{Sofikitis2017,Sofikitis2018,Kannis2021,Baylor2023,Wirtz2023}.

High-fidelity integrated modeling and hardware testing will be essential to address these issues. Additionally, because SPF’s enhancement of net electric power can mitigate the trade-off between fusion power and tritium burn efficiency~\cite{Abdou2021,Whyte2023,Parisi_2024f}, it could be instrumental in achieving tritium self-sufficiency \cite{Meschini2023}. On the economic front, using SPF to increase net electric output by several multiples could dramatically lower the capital cost per unit of electric generation capacity~\cite{Schwartz2023}. This would improve the competitiveness of early-generation fusion plants even if detailed cost analyses remain to be performed.

While this study has focused on tokamak and stellarator-like FPPs, other fusion concepts such as  inertial confinement~\cite{More1983,Goel1988,Nakao1992,Temporal_2012,AbuShawareb2022,Hu2023,Mehlhorn2024}---may also benefit from SPF, particularly systems that are close to engineering breakeven. Complementary to efforts aimed at improving plasma confinement~\cite{Wagner1982,Galambos1995,Stambaugh1998,Oyama2009,Kaye2021,Lunsford2021,Ding2024}, spin-polarized fuel presents a promising path toward the accelerated deployment of first-generation fusion power, motivating further SPF research~\cite{Garcia2023,Baylor2023,Heidbrink2024,Garcia2025}.

\begin{table*}[bt!]
\centering
\begin{tabular}{|c|c|B|c|B|c|B|c|B|c|B|c|B|}
\hline
& \textbf{STEP1} & \textbf{STEP1-P} & \textbf{STEP2} & \textbf{STEP2-P} & \textbf{\shortstack{ARC\\-Class}} & \textbf{\shortstack{ARC\\-Class-P}} & \textbf{\shortstack{ITER+\\Li6}} & \textbf{\shortstack{ITER+\\Li6-P}} & \textbf{ITER} & \textbf{ITER-P} & \textbf{\shortstack{ARIES-\\CS}} & \textbf{\shortstack{ARIES-\\CS-P}} \\
\hline
$P_\mathrm{fus}$/MW    & 1800 & 3420 & 1600 & 3040 & 525  & 998  & 500  & 950  & 500  & 950  & 2440  & 4636 \\
$P_{\mathrm{Li}6}$/MW  & 540  & 1026 & 480  & 912  & 183  & 348  & 174  & 331  & 0    & 0    & 480   & 912  \\
$P_\mathrm{e}$/MW      & 100  & 942  & 250  & 999  & 190  & 445  & 20   & 262  & -50  & 130  & 1000  & 2130 \\
$P_\mathrm{recirc}$/MW & 836  & 836  & 582  & 582  & 93   & 93   & 250  & 250  & 250  & 250  & 256   & 256  \\
$\eta$                 & 0.4  & 0.4  & 0.4  & 0.4  & 0.4  & 0.4  & 0.4  & 0.4  & 0.4  & 0.4  & 0.43  & 0.43 \\
TBR                    & 1.1  & 1.1  & 1.1  & 1.1  & 1.1  & 1.1  & 1.1  & 1.1  & 0    & 0    & -     & -    \\
$Q^\mathrm{eng}$       & 1.12 & 2.13 & 1.43 & 2.72 & 3.05 & 5.78 & 1.07 & 2.05 & 0.80 & 1.52 & 4.90  & 9.32 \\
$\mathcal{P}$          & 1    & 1.9  & 1    & 1.9  & 1    & 1.9  & 1    & 1.9  & 1    & 1.9  & 1     & 1.9  \\
$\mathcal{G}$          & -    & 9.42 & -    & 4.00 & -    & 2.34 & -    & 13.1 & -    & -    & -     & 2.13 \\
$\mathcal{G}/\mathcal{P}$ & - & 4.96 & -    & 2.11 & -    & 1.23 & -    & 6.89 & -    & -    & -     & 1.12 \\
\hline
\end{tabular}
\caption{Key parameters for the base (unpolarized) and polarized FPP scenarios discussed. STEP1, STEP2, ARC, ITER+Li6, ITER, and ARIES-CS are each examined with ($\mathcal{P}>1$) and without ($\mathcal{P}=1$) spin-polarized fuel.}
\label{table:PowerCases}
\end{table*}

\section{Acknowledgments}

We are grateful for conversations with A. Rutkowski, A.M. Sandorfi, and J.A. Schwartz. This work was supported by the U.S. Department of Energy under contract numbers DE-AC02-09CH11466, DE-SC0022270, DE-SC0022272. The United States Government retains a non-exclusive, paid-up, irrevocable, world-wide license to publish or reproduce the published form of this manuscript, or allow others to do so, for United States Government purposes.

\section{Data Availability Statement}

The data that support the findings of this study are available from the corresponding author upon reasonable request.

\appendix

\section{Engineering Triple Product Derivation} \label{app:triple_product_engineering}

The total fusion power in a volume $V$ is
\begin{equation}
P_{\mathrm{fus}} = \mathcal{P}\,\frac{n^2}{4}\,\langle\sigma v\rangle\,E_{\mathrm{fus}} V,
\end{equation}
Neutron capture in a lithium blanket contributes additional thermal power. Defining
\begin{equation}
\chi \equiv 1 + \mathrm{TBR}\,\frac{E_{\mathrm{Li}6}}{E_{\mathrm{fus}}},
\end{equation}
the gross thermal power available is
\begin{equation}
P_\mathrm{th} = \chi P_{\mathrm{fus}},
\end{equation}
which is converted to electricity with an overall thermodynamic efficiency \(\eta\), yielding a gross electric power of
\begin{equation}
P_\mathrm{e,gross} = \eta \chi P_{\mathrm{fus}}.
\end{equation}
In the plasma, the energy loss rate is approximated by
\begin{equation}
P_{\mathrm{loss}} = \frac{3nT}{\tau_E} V + P_\mathrm{B},
\end{equation}
where \(T\) is the plasma temperature, \(\tau_E\) is the energy confinement time, and $P_\mathrm{B}$ is the Bremsstrahlung power for a hydrogen plasma
\begin{equation}
    P_\mathrm{B} = C_\mathrm{B} n n_e T_e^{1/2} V.
\end{equation}
Here $n$ is total deuterium and tritium density, $n_e$ is the electron density, and $C_\mathrm{B}$ is the constant $3.34 \times 10^{-21}$ keV $\mathrm{m}^3$/s.
A fraction \(f\) of the fusion power is deposited as \(\alpha\)-particle heating:
\begin{equation}
P_{\alpha} = f\,P_{\mathrm{fus}}.
\end{equation}
If \(P_{\alpha}\) is insufficient to cover \(P_{\mathrm{loss}}\), additional external heating \(P_{\mathrm{ext}}\) is required. We assume that all external heating is provided via the reactor’s own recirculating power. The total recirculating power is
\begin{equation}
P_{\mathrm{recirc}} = \beta\,P_{\mathrm{fus}},
\end{equation}
where \(\beta\) accounts for both auxiliary loads (e.g., pumps, magnets) and the external heating \(P_{\mathrm{ext}}\). The net electric power available is then
\begin{equation}
P_{\mathrm{e}} = \eta \chi P_{\mathrm{fus}} - P_{\mathrm{recirc}},
\end{equation}
and engineering breakeven requires
\begin{equation}
P_{\mathrm{e}} \ge 0 \quad\Longrightarrow\quad \eta \chi P_{\mathrm{fus}} \ge \beta\,P_{\mathrm{fus}}.
\end{equation}
giving
\begin{equation}
\eta\,\chi \ge \beta.
\end{equation}

To sustain the plasma the power loss must be met by the internal \(\alpha\)-heating plus the externally supplied heating (which is drawn from \(P_{\mathrm{recirc}}\)). Thus we require
\begin{equation}
P_{\mathrm{loss}} \le P_{\alpha} + P_{\mathrm{ext}}.
\label{eq:Ploss}
\end{equation}
The heating power is a fraction $\delta$ of all recirculating power,
\begin{equation}
    P_\mathrm{B} + P_{\mathrm{ext}} = \delta P_\mathrm{recirc} = \delta \beta P_\mathrm{fus}.
\end{equation}
At engineering breakeven, $\eta \chi = \beta$, and therefore at engineering breakeven the heating power satisfies
\begin{equation}
    P_\mathrm{B} + P_{\mathrm{ext}} \bigg{|}_{Q^\mathrm{eng}} = \delta \eta \chi P_\mathrm{fus}.
\end{equation}
Thus, at breakeven \Cref{eq:Ploss} satisfies
\begin{equation}
 P_\mathrm{B} + \frac{3nT}{\tau_E} \bigg{|}_{Q^\mathrm{eng} } = (f+\delta \eta \chi) P_{\mathrm{fus}},
\end{equation}
where we substituted \(P_{\alpha} = f\,P_{\mathrm{fus}}\). Replacing \(P_{\mathrm{fus}}\) and $P_\mathrm{B}$ by their expressions and some further algebraic manipulations gives the triple product
\begin{equation}
nT\tau_E \bigg{|}_{Q^\mathrm{eng}} = \frac{12\,T^2}{(f+\delta \eta \chi) \mathcal{P} \langle\sigma v\rangle E_{\mathrm{fus}} - 4 C_\mathrm{B} T^{1/2}},
\label{eq:tripleproduct_engineering_appendix}
\end{equation}
where we assumed $n_i = n_e$ and $T_i = T_e$. Thus, by including the external heating \(P_{\mathrm{ext}}\) in the recirculating power \(P_{\mathrm{recirc}} = \beta\,P_{\mathrm{fus}}\), we obtain the above triple product condition for engineering breakeven. In the limit $\delta \to 0$, \Cref{eq:tripleproduct_engineering_appendix} reduces to the Lawson criterion for scientific ignition where $Q^\mathrm{sci} \equiv P_\mathrm{fus} / P_\mathrm{ext} \to \infty$ \cite{Lawson1957,Wurzel2022}.

\bibliography{EverythingPlasmaBib} %

\end{document}